\documentclass[aps,nopacs,amsmath,amssymb,twocolumn]{revtex4}
\usepackage{graphicx}
\usepackage{epstopdf}

\usepackage{tikz}
\usetikzlibrary{decorations.pathreplacing}
\usepackage{subcaption}

\def\beq{\begin{eqnarray}}
\def\eeq{\end{eqnarray}}
\def\be{\begin{equation}}
\def\bel{\begin{equation}\label}
\def\beel{\begin{eqnarray}\label}
\def\ee{\end{equation}}

\def\bm{\begin{math}}
\def\me{\end{math}}

\def\q{\quad}
\def\qq{\qquad}

\newcommand{\Fig}[1]{Fig.~\textup{\ref{#1}}}
\newcommand \nn {\nonumber}
\newcommand \bei {\begin{itemize}}
\newcommand \eei  {\end{itemize}}

\begin{document}
\bibliographystyle{apsrev}

\title{Anomalous  aggregation regimes of temperature-dependent Smoluchowski equations
}

\author{A.I.~Osinsky$^{1}$}
\author{N. V. Brilliantov$^{1,2}$}
\affiliation{$^{1}$Skolkovo Institute of Science and Technology, Moscow, Russia}
\affiliation{$^{2}$Department of Mathematics, University of Leicester, Leicester LE1 7RH, United Kingdom}

\begin{abstract}
Temperature-dependent Smoluchowski equations describe the ballistic agglomeration. In contrast to the standard Smoluchowski equations for the evolution of cluster densities with constant rate coefficients, the temperature-dependent equations describe both -- the evolution of the densities as well as cluster temperatures, which determine the aggregation rates. To solve these equations, we develop a novel Monte Carlo technique based on the low-rank approximation for the aggregation kernel. Using this highly effective approach, we perform a comprehensive study of the phase diagram of the system and reveal a few surprising regimes, including permanent temperature growth and ``density separation'', with a large gap in the size distribution for middle-size clusters. We perform classification of the aggregation kernels for the temperature-dependent equations and conjecture the lack of gelation. The results of our scaling analysis agree well with the simulation data. 

\end{abstract}

\maketitle


{\it Introduction.} Aggregation processes are very ubiquitous in nature at different time and space scales, e.g.  \cite{BrilliantovPNAS2015, Erik17, BlumErosion2011, Falkovich2002, Igor14,smoluchowski-deu,Leyvraz2003,krapbook,Das,Mazza}. The classical tool to describe the aggregation kinetics is the celebrated Smoluchowski equations \cite{smoluchowski-deu}, which deal with the density of aggregates $n_k(t)$. Here the subscript $k$ specifies the size of the aggregate, comprised of $k$ monomers -- the elementary units;  these equations read for $1\leq k \leq \infty$ \cite{Leyvraz2003,krapbook}: 
\begin{equation}\label{n-eq}
\begin{aligned}
  & \frac{d}{{dt}}{n_k} = \frac{1}{2}\sum\limits_{i + j = k} {{C_{ij} }{n_i}{n_j}}  - \sum\limits_{j = 1}^\infty  {{C_{kj} }{n_k}{n_j}} , \\
\end{aligned}
\end{equation}
The kinetic coefficients $C_{ij}$ quantify the reaction rates between the aggregates of size $i$ and $j$. They may either follow from a microscopic model or be constructed as empirical expressions \cite{Leyvraz2003,krapbook}. Importantly, the classical Smoluchowski theory treats these coefficients as time-independent so that the infinite set \eqref{n-eq} forms a closed system of equations. 

It has been recently shown that for the ballistic agglomeration, the rate coefficients are time-dependent,  $C_{ij}=C_{ij}(t)$, as they are functions of the energy density (temperature $T$) of the system \cite{Mazza,BOK2020} or of partial energy densities associated with the aggregates of size $i$ (partial temperatures $T_i$) \cite{BFP2018}. In the course of time, these quantities vary. Hence to make the system of equations closed, one needs to supplement  Eqs. \eqref{n-eq} for densities by the set of equations for temperatures $T_k(t)$ \cite{BFP2018}: 
\begin{equation}\label{nt-eq}
\frac{d}{{dt}}{n_k}{\theta _k} = \frac{1}{2}\sum\limits_{i + j = k} {{B_{ij}}{n_i}{n_j}}  - \sum\limits_{j = 1}^\infty  {{D_{kj}}{n_k}{n_j}} . 
\end{equation}
Here $\theta_k =T_k/m_k$ and $m_k=m_1k$ is the mass of  aggregates of size $k$, with the diameter $\sigma_k =\sigma_1 k^{1/3}$.  Eqs. \eqref{n-eq} and \eqref{nt-eq} form a closed set; all of the rate coefficients $C_{ij}$,  $B_{ij}$ and $D_{ij}$ depend on the partial temperatures $T_i(t)$ and $T_j(t)$, see the Supplementary Material (SM). Moreover a microscopic analysis of particles collisions shows that besides of temperatures, the reaction rates sensitively depend on the interaction potential between particles, which may be put, for a wide class of interactions, into the form: 

\be\label{W}
W_{ij} = a \left(i^{1/3}j^{1/3}\right)^{\lambda _1} \left(i^{1/3} + j^{1/3} \right)^{-\lambda _2},
%
%
\ee 
where the constant $a$ specifies the interaction energy, while $\lambda_{1} $ and $\lambda_{2}$ quantify the dependence of $W_{ij}$ on the size of particles $i$ and $j$. For instance, $\lambda_1=\lambda_2=4/3$ corresponds to the adhesive surface interactions,  $\lambda_1=\lambda_2=3$ stands for the dipole-dipole interactions  and $\lambda_1=3$, $\lambda_2=1$ refers to the gravitational or Coulomb interaction, when the particles charges scale as their masses \cite{BFP2018}. Still, the aggregation rate is determined not directly by $W_{ij}$, but by its ratio to the characteristic kinetic energy of  the colliding particles, that is, by the dimensionless quantity, 

\be\label{qij}
{q_{ij}}  = \frac{{{W_{ij}}}}{{{\varepsilon ^2} \mu_{ij} \left( {{\theta _i} + {\theta _j}} \right)}}, \ee
where $\mu_{ij}= m_i m_j /(m_i+m_j)$ and $\varepsilon$ is the restitution coefficient;  it quantifies the dissipative losses at particles collisions. Hence $C_{ij}(t)= C_{ij}\left(q_{ij}(t) \right)$ and similarly $B_{ij} $ and $D_{ij}$, see SM for details. 

{\it Temperature-dependent  Monte Carlo.} The classical Smoluchowski system \eqref{n-eq} may be analytically solved only for  a few kernels $C_{ij}$ \cite{bilinearkernel, Leyvraz2003,krapbook}. Generally, however, it requires a numerical analysis, e.g. \cite{compfvs, Immanuel2003, chaud2014, matveev2015, BallOscil, colmPRE2018}.  Still, even the numerical solution is rather challenging, as the system of equations is infinite. The solution of the complete set \eqref{n-eq} and \eqref{nt-eq} brings  further complications. 
To address this problem, we develop a novel temperature-dependent Monte Carlo (MC) method, which is extremely efficient. It allows to investigate the behavior of huge systems in a wide range of parameters, which was not possible with the previous methods \cite{gillespie, inverse, monte-random, majorant, SabelfeldEremeev}. The main idea of the method is to exploit the low-rank approximation for the kinetic kernels, which has been successfully applied to solve classical Smoluchowski equations \cite{MKSTB}. We adopt this approach to the MC scheme, with the extension for the temperature dependence, that is, for Eqs. \eqref{n-eq} and \eqref{nt-eq}. The method explicitly recalculates temperatures for each aggregate size without generating particle velocities. For our low-rank MC simulations only $O \left(r \log M \right)$ operations are needed for each collision. For the system rank $r \equiv 3$, it allows performing $\sim 10^6$ collisions every second without any use of parallel computation. Here $M$ is the maximum cluster mass, and $r$ is the rank of the kernel approximation. The implementation and detail of the new low-rank MC approach to our problem is discussed in SM. 


{\it Numerically-obtained phase diagram.} We observe that the system obeying the  temperature-dependent Smoluchiowski equations demonstrates extremely rich behavior: various temperature and aggregation regimes, including regimes of temporal and permanent temperature growth, the so-called ``separation'' regime, as well as classical aggregation with cooling. We vary two main parameters: $\Lambda=\lambda_1= \lambda_2$, which specifies, how the kinetic rates depend on the aggregates' size and $q=q_{11}(0)=\frac{a}{T_1(0)} \frac{1}{\varepsilon^2 2^{\Lambda}}$, which quantifies   the (initial) ratio of the potential and kinetic energy of monomers.  In simulations we use mono-disperse initial conditions (only monomers are available at $t=0$) with the initial dimensionless density $n_1(0) = 0.0955$ 
and temperature $T_1(0)=1$. We also use $\varepsilon=0.99$,  $m_1=1$ and $\sigma_1=1$. 
\begin{figure}[ht]
\includegraphics[width=\columnwidth]{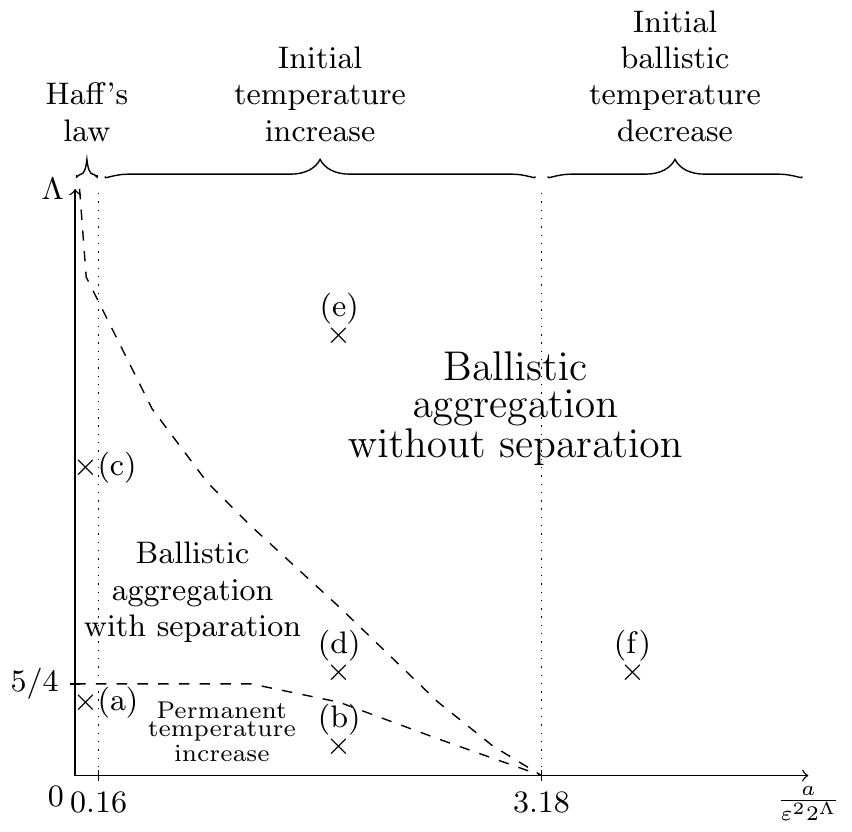}
\caption{
Phase diagram for the temperature-dependent aggregation, Eqs. \eqref{n-eq} and \eqref{nt-eq}. The dashed lines demarcate different kinetic regimes. They are obtained with the simulation grid size of about $0.6$ for both coordinates; close to the borders the grid step reduced to $0.075$ for $q$ and to $0.2$ for $\Lambda$. Temperature dependencies for the phase points indicated by crosses are depicted in \Fig{typ-fig}.
}
\label{diag-fig}
\end{figure}

\begin{figure}[ht]
\begin{subfigure}[b]{0.49\columnwidth}
\centering
\includegraphics[width=\columnwidth]{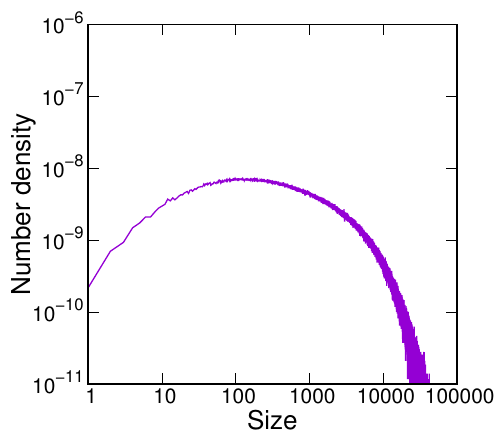}
\end{subfigure}
\begin{subfigure}[b]{0.49\columnwidth}
\centering
\includegraphics[width=\columnwidth]{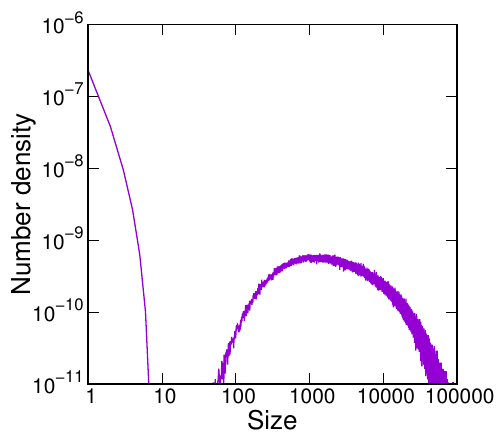}
\end{subfigure}
\caption{Cluster size distribution $n_k$ as a function of $k$ at $t = 10000$. Left panel: Size distribution without separation ($\Lambda=1.4$, $q=3.8$). Right panel: Size distribution with separation -- density of middle-size clusters is almost vanishing ($\Lambda=1.4$, $q=1.8$). 
}
\label{sep-fig}
\end{figure}

The phase diagram in \Fig{diag-fig}  illustrates the areas in the parametric space $(q,\Lambda)$ corresponding to different evolution regimes. The most surprising is the aggregation ``with separation'' and aggregation with permanent temperature growth. In the former case, the cluster size distribution demonstrates an impressive density gap between small and large aggregates. That is, the density of intermediate-size clusters can be several orders of magnitude lower than that of small and large clusters, see Fig. \ref{sep-fig}. This may be explained by very large reaction rates for these clusters. In the latter case, the aggregation takes place in such a way that the rate of energy loss due to the agglomeration is lower than the aggregation rate. This results in the increasing energy per particle, that is, in growing temperature. Next, one can classify the regimes by evolution of temperature -- it may follow the Haff's law with a continuous decay of temperature, or may alter the temperature regime, from the decay to growth and then back to decay; finally, a permanent growth from the very beginning is also possible. 

\begin{figure}[p]

\begin{subfigure}[t]{\columnwidth}
\centering
\includegraphics[width=\columnwidth]{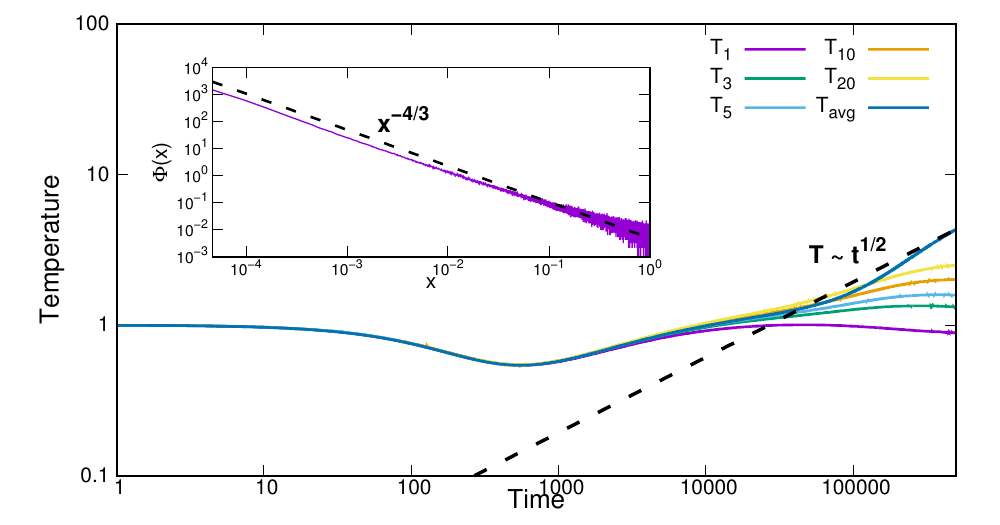}
\caption{$\Lambda = 1$, $q_{11}(0) = 0.075$.
}
\label{Figa}
\end{subfigure}
\begin{subfigure}[t]{\columnwidth}
\centering
\includegraphics[width=\columnwidth]{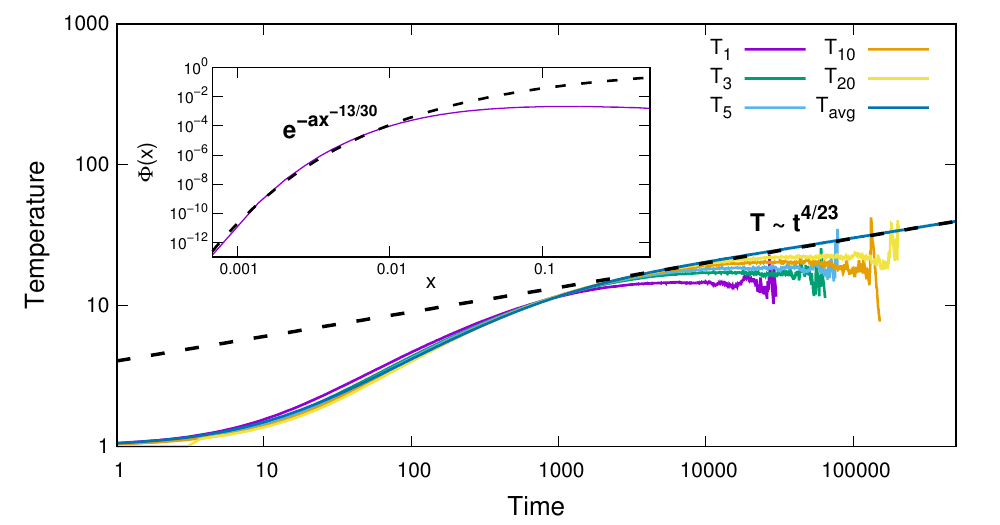}
\caption{$\Lambda = 0.4$, $q_{11}(0) = 1.8$.}
\label{Figb}
\end{subfigure}

\begin{subfigure}[t]{0.49\columnwidth}
\centering
\includegraphics[width=\columnwidth]{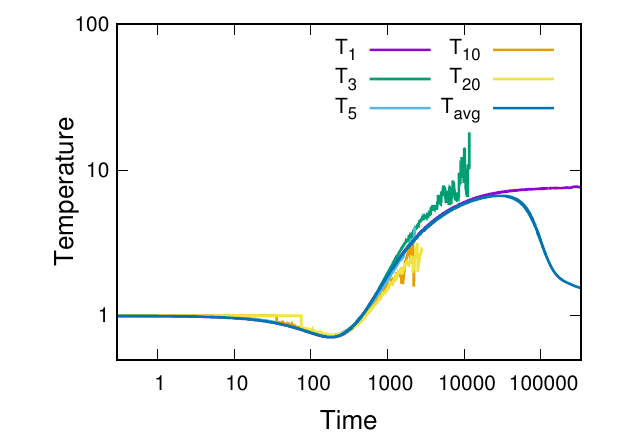}
\caption{$\Lambda = 4.2$, $q_{11}(0) = 0.075$.
}
\end{subfigure}
\begin{subfigure}[t]{0.49\columnwidth}
\centering
\includegraphics[width=\columnwidth]{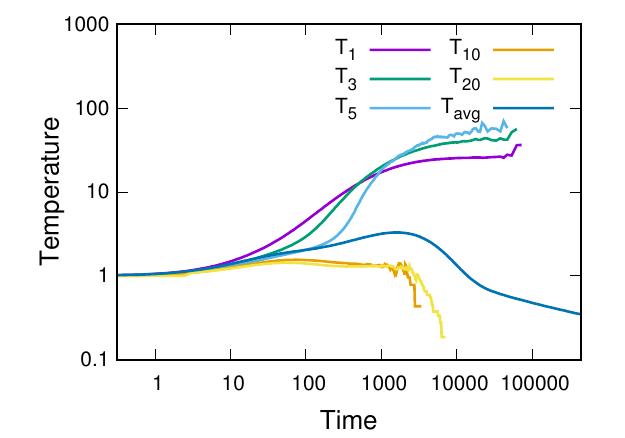}
\caption{$\Lambda = 1.4$, $q_{11}(0) = 1.8$.}
\end{subfigure}
\begin{subfigure}[t]{0.49\columnwidth}
\centering
\includegraphics[width=\columnwidth]{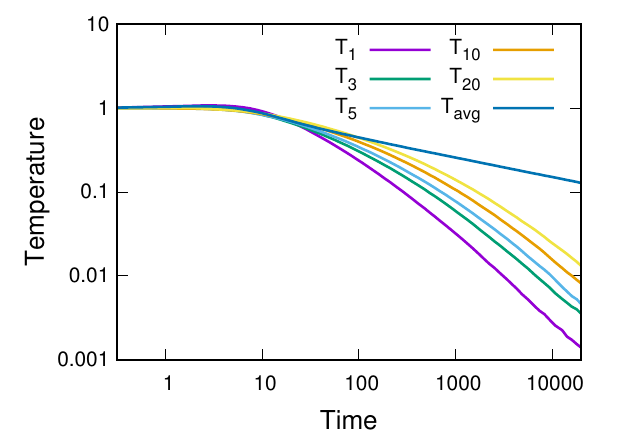}
\caption{$\Lambda = 6$, $q_{11}(0) = 1.8$.}
\end{subfigure}
\begin{subfigure}[t]{0.49\columnwidth}
\centering
\includegraphics[width=\columnwidth]{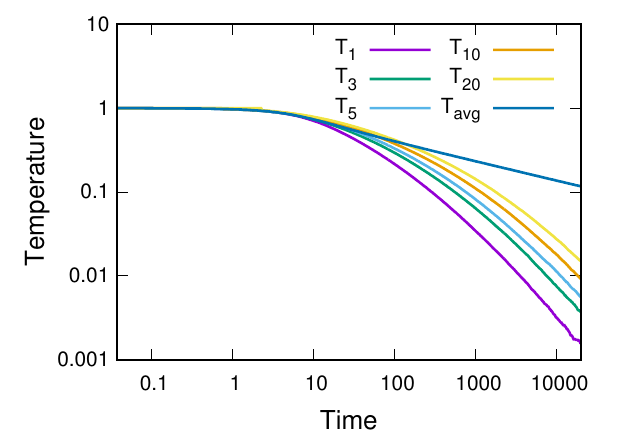}
\caption{$\Lambda = 1.4$, $q_{11}(0) = 3.8$.
}
\end{subfigure}
\caption{Typical examples of temperature evolution, corresponding to different phase points in \Fig{diag-fig}. MC simulations are performed with $10^7$ particles. Period of temperature increase exists everywhere, except panel (f); for  the panel (e) $T_{\max} \approx 1.05$. Insets in panels (a) and (b) depict the scaling function $\Phi(x)$ for the Class I (a) and Class III (b) kernels; the dashed lines -- the scaling theory, see the text for detail.
}
\label{typ-fig}
\end{figure}

Fig. \ref{typ-fig} illustrates the typical examples of temperature evolution corresponding to the different regions of the phase diagram in Fig. \ref{diag-fig}. The figure shows  all the mentioned scenarios realized on the $(q, \Lambda)$ plane. The set of equations is very complicated for a whole analytical study; below we present a qualitative analysis of the system behavior. 


{\it Initial behavior.} We start from classifying the initial behavior. For the qualitative analysis we use the average temperature $T$ and ignore the difference of partial temperatures, $T_i=T$ and $\theta_i = T/i$ (recall that $m_i=i$). Then the densities $n_k$ follow Eqs. \eqref{n-eq}, while the equation for the average temperature reads,  
\bel{T}
\frac{d}{{dt}}nT =  - \sum\limits_{i,j = 1}^\infty  {{P_{ij}}{n_i}{n_j}}.  \ee
Here $n(t)=\sum_{i=1}^{\infty}n_i(t)$ is the total density and the rate coefficients simplify to (see \cite{BFP2018}): 
\beel{CP}
  && C_{ij} = 2\sqrt{2 \pi T} \sigma_{ij}^2 \sqrt{i^{-1} +j^{-1} } (1 - f_{ij}), \\
  &&P_{ij}\! =\! \frac{4}{3} \sqrt{2 \pi T^3} \sigma_{ij}^2  \sqrt{i^{-1} +j^{-1} }  
  \left(\! 1 \!- \!g_{ij} \!+ \!\frac12 \!\left( 1 \!- \!\varepsilon^2 \right) g_{ij} \!\right), \nonumber \\
  &&f_{ij} = e^{-q_{ij}} (1 + q_{ij}), \qq \q q_{ij} = {W_{ij}}/({\varepsilon^2 T}), \nonumber \\
  &&g_{ij} = e^{-q_{ij}} (1 + q_{ij} + q_{ij}^2/2),   \nn \eeq
where $\sigma_{ij} =\frac12 (i^{1/3} +j^{1/3})$. Initially monomers strongly dominate, that is, $n \approx n_1$, which yields from Eqs. \eqref{n-eq} and \eqref{T}: 

\bel{nT}
\dot n \approx  - \frac{1}{2}{n^2}{C_{11}}, \qq \q  
  \dot nT + n\dot T \approx  - {n^2}{P_{11}}. \ee
From the last equations and Eqs. \eqref{CP} for $C_{11}$ and $P_{11}$ follows, 
\bel{Tini}
\dot T  \approx  - n{\sigma_{11}^2}\sqrt {4\pi {T^3}} \left( {\frac{1}{3} + f_{11} - \frac{4}{3}{g_{11}} 
+ \frac{2}{3}\left( {1 - {\varepsilon ^2}} \right){g_{11}}} \right). \ee
Three different scenarios may be realized.  \\
{\it (a)} Small $q$ (high initial temperatures),   $q^2 \ll 1-\varepsilon^2$:
$$
\dot T \approx - 2n{\sigma_{11}^2}\sqrt {4\pi {T^3}} \left( {1 - {\varepsilon ^2}} \right)/3 \q \to \q \dot T \sim -T^{3/2}, 
$$
where we take into account that in this case $\dot n(0) \ll \dot T(0)$. This regime corresponds to the Haff's law of non-aggregative cooling, 
\bel{Haff} T \approx (1+ct)^{-2} , \ee 
where $c$ is a constant \cite{BFP2018}. \\
{\it (b)} For $1-\varepsilon^2 \ll q \ll 1$ we find: 
$$
\dot T \approx n {\sigma_{11}^2}\sqrt {4\pi {T^3}} q_{11}^2 / 2 = - T \dot n / n,
$$
that is, $T\approx n^{-1}$. The solution to the last equation reads,  
\bel{mid_q}
T = (1 + c t)^{2/5}, \qq \q n = (1 + c t)^{-2/5} , \ee
where we also use Eqs. \eqref{nT} and \eqref{CP}. The transition from the initial decay of temperature, as in \eqref{Haff},  to the initial growth, as in \eqref{mid_q}, happens when $\dot{T}(0)=0$, yielding
\bel{q_dem1}
q_1 \approx 2\sqrt{(1-\varepsilon^2)/3} \approx 0.16, \ee
for $\varepsilon =0.99$, in agreement with the simulations, Fig. \ref{diag-fig}.\\
{\it (c)} Large $q$ (low initial temperature), $e^{-q} \left(1 + q) \right) \ll 1$, give rise to the equation, 
$$
  \dot T \approx - \frac{1}{3} n {\sigma_{11}^2}\sqrt {4\pi {T^3}} = \frac{1}{3} T \dot n / n,  
$$
and $T = n^{1/3}$. This, together with Eqs. \eqref{nT} and \eqref{CP} leads to the  solution:
\bel{large_q} 
  T = (1 + c t)^{-2/7}, \qq \q n = (1 + c t)^{-6/7}. \ee
Unfortunately, it is not possible to observe this behavior for a sufficiently long time. Nevertheless, Eq. \eqref{large_q} shows that the temperature starts to decrease immediately. The value of $q$, demarcating the initial increase and decrease of temperature 
follows from $\dot{T}(0)=0$, applied to Eq. \eqref{Tini}:   
$$
e^{q} - 1 - q - 2q^2 + 2 (1 - \varepsilon^2) (1 + q + q^2/2) \approx 0.
$$
The result reads, 
\bel{q_dem}
q_2 \approx 3.18, \ee
again, in agreement with the simulation data, see Fig. \ref{diag-fig}. 


{\it Scaling regimes and ``separation''.} After a transient period when large clusters emerge,   an  aggregating system often enters a scaling regime, where the cluster size distribution may be described by a scaling function, $n_k(t)=s^{-2}\Phi(k/s)$. Here the typical cluster size  $s = M_2(t)/M_1$ ($M_l = \sum_k n_k k^l$) is large, $s\gg 1$.  It scales as a power-law,  $s \sim t^{z}$,  for  $t \to \infty$  \cite{Leyvraz2003,krapbook}. 
For temperature-dependent aggregation  an additional temperature scaling $T\sim t^{-\beta}$ may emerge.  The scaling exponents $z$ and $\beta$ are  determined by the characteristic values $q_{ij}$ for $i,j \sim s$ and $\Lambda$. 

For small $q_{ss}\ll 1 $ (high temperature) the rates $C_{ij}$ are homogeneous functions of $i,j$; the expansion of these coefficients \eqref{CP} yields, 
\beel{Cijexp}
C_{ai,aj} &\simeq &  a^{\mu} C_{ij};  \qq \qq  \mu = \frac{1}{6} + \frac{2}{3} \Lambda, \\
\label{Cijexp1}
C_{bi,j} & \simeq & b^{\nu} C_{ij} \qq \qq  ~~ \nu = \frac{2}{3}, \eeq  
for $b \gg 1$. Similar expressions may be found  for the rates $P_{ij}$, yielding the following exponents, reported in \cite{BFP2018}: 
\bel{zbet}
 z = \frac{6}{5 - \Lambda}, \quad \qq \beta = -\frac{2 \Lambda}{5 - \Lambda}.\ee 
This surprising behavior with the increasing average temperature $T$ has been first reported in Ref. \cite{BFP2018} as an intermediate regime for a finite-size system. Here we confirm it for different values of $\Lambda$ (see Fig. \ref{typ-fig}) and demonstrate that it is permanent and stable in the thermodynamic limit, in agreement with the findings of \cite{meode}. The permanent  growth of $T$ is observed for  $\Lambda \leq 5/4$, see Fig. \ref{diag-fig}.  

When $\Lambda > 5/4$, which corresponds to $\mu >1$, the classical criterion for gelation is fulfilled \cite{van-dongen,Leyvraz2003}. In this case, all mass is accumulated in a single gigantic gel, adsorbing all particles; the cluster densities become zero.  For our system the situation is more involved -- for very large clusters the condition $q_{ij} \ll 1 $ is violated and converts into the opposite one, $q_{ij} \gg 1$, that is $f_{ij} \simeq 0$, see Eqs. \eqref{CP}. The latter implies a purely ballistic agglomeration: 
\bel{baag}
C_{ij} \simeq  2\sqrt{2 \pi T} \sigma_{ij}^2 \sqrt{i^{-1} +j^{-1} }; \qq C_{ai,aj}=a^{\mu_b}C_{ij}, \ee
with $\mu_b=1/6 <1$, which is strictly non-gelling. Hence the appearance of the pronounced gap in the cluster size distribution for the middle-size aggregates (we call it ``separation'') is the result of the ``quasi-gelation'' process. For small clusters, the scaling analysis is not applicable. The lack of "true"gelling in these systems is argued below. 

Large  $q_{ss}$ (low temperature) with $e^{-q_{ss}}(1+q_{ss}) \ll 1$ imply aggregation with cooling \cite{BFP2018}: 
\bel{zbet1}
 z = 1, \quad \qq \beta = 1/3 .\ee 
In simulations we observe, however, different exponent, $T \sim t^{-0.22}$, which is caused by the failure of the approximation $T_i=T$  in this limit, see  \Fig{typ-fig} (e), (f). 


{\it The lack of gelation.} Qualitatively, the lack of gelation for temperature-dependent Smuluchowski equations follows from the conversion of the aggregation rates $C_{ij}$ for very large clusters into the ballistic, non-gelling case. These arguments exploit, however, the gelling criterion for the classical case. Hence it is desirable to prove the lack of gelation directly for the complete set of equations, \eqref{n-eq} and \eqref{nt-eq}. To this end, we need to prove that the moments of the density distribution would always be bounded \cite{van-dongen}. Indeed, when the gelation happens, the second moment $M_2(t) = \sum_k n_k k^2$ becomes infinite \cite{van-dongen}.  To estimate $M_2(t)$  we also need the second moment for temperatures $M_{2, \theta}(t) = \sum_k n_k \theta_k k^2$. Multiplying Eqs. \eqref{n-eq} and \eqref{nt-eq} with  $k^2$ and summing them over all $k$ we arrive at (see SM for the derivation): 
\begin{equation}
\label{eq:secondmomsys}
\begin{aligned}
  \frac{d M_2}{dt} & \leqslant C_1 M_2 + C_2 \sqrt{M_2 M_{2, \theta}}, \\
  \frac{d M_{2, \theta}}{dt} & \leqslant C_3 \sqrt{M_2 M_{2, \theta}} \\
  & - \sum\limits_{i,j} \left( T_i - T_j \right) \left( i - j \right) E_{ij} n_i n_j.
\end{aligned}
\end{equation}
Here $E_{ij}$ is the symmetric positive function, which describes the rate of energy exchange in bouncing collisions (see SM). For non-aggregating granular mixtures it was theoretically and numerically shown that $T_i$ grows with the size of the aggregates $i$, for all distribution with a density dominance of monomers \cite{bodrova2014}; we apply this for the studied aggregating systems. In all our simulations we always observed that either $T_i = T_j$, or $T_i>T_j$ for $i>j$ (except for a period of time with separation). Hence the conjecture $T_i>T_j$ for $i>j$ is justified. In this case the last term in Eqs. \eqref{eq:secondmomsys} is negative, yielding the result, 
\bel{M2lim}
M_2(t) \leqslant e^{\left( C_1 + C_2 + C_3 \right) t} ,\ee
that is, the moment $M_2(t)$ is bounded at any time instant and gelation is not possible. 


{\it Classification of the rate kernels.} In the classical temperature-independent theory, the aggregation rate kernels $C_{ij}$ are classified to be of three types, corresponding to qualitatively different behavior \cite{van-dongen,Leyvraz2003}. The temperature-dependent theory possesses an extended set of kernels: $C_{ij}$, $B_{ij}$ and $D_{ij}$. For practical reasons, it would be worth to have the according classification aligned with the classical one.

For small $q \ll 1$ one can apply the expansion leading to Eqs. \eqref{Cijexp} and \eqref{Cijexp1}. From these equations follows that Class I kernel is realized for $\Lambda > 3/4$, which corresponds to $\nu < \mu$. This class is characterized by the power-law size dependence of the density   of small clusters, $\Phi(x) \sim x^{-(1+\mu - \Lambda/2)}$ for $x\ll 1$ (\Fig{Figa}), where $T(x) \sim x^{\Lambda/3}$ for small $x$. The case $\Lambda =  3/4$ corresponding  to $\nu = \mu$ is classified as  Class II kernels. It is  characterized by similar power-law dependence for small clusters,  $\Phi(x) \sim x^{-\tau }$, as for the Class I, but the exponent $\tau$ is not universal. Finally, Class III is realized for $\Lambda < 3/4$, that is for $\nu > \mu$. It is characterized by the exponential disappearance of small clusters, $\Phi(x) \sim  \exp(-ax^{\mu-\nu-\Lambda/2})$ for $x\ll 1$ (\Fig{Figb}) \cite{van-dongen,Leyvraz2003}.

In the opposite case of large $q \gg 1$ temperature always decreases, $f_{ij} \simeq 0$ and kernels tend to the ballistic one, given by Eq. \eqref{baag}, with $\mu_b=1/6$. This regime is stable and again corresponds to the Class III kernel. Note that in aggregation with separation small-size clusters  eventually disappear, leading to the same scaling solution as for $q \gg 1$. Therefore, all systems locating on the phase diagram outside permanently increasing temperature region always converges to the same Class III scaling with $\mu = \mu_b$.


{\it Conclusion.} We investigate the aggregation kinetics for temperature-dependent Smoluchowski equations. In contrast to standard Smoluchowski equations, which describe the evolution of clusters densities of different sizes with time-independent rates, we consider two coupled sets of equations -- one for the cluster densities and another for the partial temperatures of the aggregates, which define the reaction rates. For the numerical solution of these sets of equations, we develop a novel, highly efficient Monte Carlo approach based on the low-rank approximation of the reaction kernels. It allows for a fast solution of huge systems of equations. We explore the system's behaviour for a wide range of parameters and obtain a complete evolution phase diagram. It possesses several surprising regimes, including the regime of permanent temperature growth and density ``separation''. We make a classification of the aggregation kernels and conjecture a lack of gelation for temperature-dependent aggregation. The results of our scaling analysis are in good agreement with the simulation results. 

\section*{Acknowledgements}

The study was supported by a grant from the Russian Science Foundation No. 21-11-00363, https://rscf.ru/project/21-11-00363/.

\appendix

\section*{SUPPLEMENTARY MATERIAL} 

Below we present the detail of the new Monte Carlo method for the solution of the temperature-dependent Smoluchowski equations. We also give the complete expressions for the rate kernels and some derivation details. References to the equations from the main text are given in bold. 

\section*{Kinetic rate coefficients for the temperature-dependent Smoluchowski equations} 
Here we write the temperature-dependent Smoluchowski equations {\bf (1)} with the explicit indication of the temperature dependence of the rate coefficients $C_{ij}$: 
\begin{equation}\label{n-eq2}
\begin{aligned}
  & \frac{d}{{dt}}{n_k} = \frac{1}{2}\sum\limits_{i + j = k} {{C_{ij} \left( T_i, T_j \right)}{n_i}{n_j}}  - \sum\limits_{j = 1}^\infty  {{C_{kj} \left( T_k, T_j \right)}{n_k}{n_j}} , \\
  & k = \overline {1,\infty }. 
\end{aligned}
\end{equation}
Here  $T_i$ and $T_j$ are temperatures of clusters of size $i$ and $j$, which generally  differ for clusters of different size.  These temperatures obey the following equations, written for the reduced variables, $\theta_i = T_i / m_i$, where $m_i =im_1$ is the mass of clusters of size $i$ \cite{BFP2018}: 

\begin{equation}\label{nt-eq2}
\frac{d}{{dt}}{n_k}{\theta _k} = \frac{1}{2}\sum\limits_{i + j = k} {{B_{ij}}{n_i}{n_j}}  - \sum\limits_{j = 1}^\infty  {{D_{kj}}{n_k}{n_j}} ,\quad k = \overline {1,\infty }.
\end{equation}
Equations {\bf (1)-(2)} have been obtained for the ballistic agglomeration of particles, interacting with the energy $W_{ij}$, Eq. {\bf (3)}, which depends on the size of the particles and on the nature of the inter-particle forces (see the main text). The according kinetic coefficients read \cite{BFP2018}:
\begin{equation}\label{sys-coef}
\begin{gathered}
  {C_{ij}} = 2\sqrt {2\pi } \sigma _{ij}^2\sqrt {{\theta _i} + {\theta _j}} \left( {1 - {f_{ij}}} \right), \hfill \\
  {B_{ij}} = 2\sqrt {2\pi } \sigma _{ij}^2\frac{1}{{\sqrt {{\theta _i} + {\theta _j}} }}\left( {{\theta _i}{\theta _j}\left( {1 - {f_{ij}}} \right)} \right. \hfill \\
  \left. {+ \frac{4}{3}{{\left( {\frac{{i{\theta _i} - j{\theta _j}}}{{i + j}}} \right)}^2}\left( {1 - {g_{ij}}} \right)} \right), \hfill \\
  {D_{ij}} = 2\sqrt {2\pi } \sigma _{ij}^2\frac{1}{{\sqrt {{\theta _i} + {\theta _j}} }}\left( {{\theta _i}{\theta _j}\left( {1 - {f_{ij}}} \right)} \right. \hfill \\
  \left. + {\frac{4}{3}\theta _i^2\left( {1 - {g_{ij}}} \right)} \right. \hfill \\
  \left. { + \frac{4\left( 1 + \varepsilon \right) j}{3 \left( i + j \right)} \left( \theta_i + \theta_j \right) \left( {{\theta _i} - \frac{\left( 1 + \varepsilon \right) j}{2 \left( i + j \right)} \left( {{\theta _i} + {\theta _j}} \right)} \right)} g_{ij} \right), \hfill \\
\end{gathered} 
\end{equation}
\begin{equation}\label{sys-coef2}
\begin{aligned}
  {f_{ij}} & = {e^{ - {q_{ij}}}}\left( {1 + {q_{ij}}} \right), \\
  {g_{ij}} & = {e^{ - {q_{ij}}}}\left( {1 + {q_{ij}} + q_{ij}^2/2} \right), \\
  {q_{ij}} & = \frac{{{W_{ij}}}}{{{\varepsilon ^2}\tfrac{{ij}}{{i + j}}\left( {{\theta _i} + {\theta _j}} \right)}}, \\
  {W_{ij}} & = a\frac{{{{\left( {{i^{1/3}}{j^{1/3}}} \right)}^{{\lambda _1}}}}}{{{{\left( {{i^{1/3}} + {j^{1/3}}} \right)}^{{\lambda _2}}}}}, \\
  {\sigma _{ij}} & = \frac{{{i^{1/3}} + {j^{1/3}}}}{2}.
\end{aligned} 
\end{equation}
Under the approximation of equal temperatures, $T_i =T$ for all $i$ (valid in some regimes), Eqs. \eqref{nt-eq2} reduce to Eqs. {\bf (5)-(6)}. 

\section*{Temperature-dependent Monte Carlo}

In order to make our simulations equivalent to the solution of temperature-dependent Smoluchowski equations, we assume the distribution of speeds to be Maxwellian for each cluster size \cite{BFP2018}. Naturally, we need a large number of clusters of each size to justify this assumption. But even if clusters of a particular size become scanty or disappear, this will not noticeably impact the overall solution, as the assumption remains true for the most of collisions.

Hence we can use for the Monte Carlo simulations the kernels $C_{ij}$, $B_{ij}$ and $D_{ij}$,  as defined in \eqref{sys-coef},  to determine the collision frequencies and the corresponding temperature variation.

The time step $\tau$ between collisions can be determined using equation \eqref{n-eq2}. Note that the factor $1 - f_{ij}$ in the kernel $C_{ij}$  defines the aggregation probability. Therefore,  to determine the time between any (not only aggregative) collisions, we can use the kernel $\hat C_{ij} = C_{ij} / \left( 1 - f_{ij} \right)$. Let us define the system volume as $V = N / n = N(t = 0) / n(t = 0)$, where $N$ is the total number of clusters and $n$ is the total cluster density. If only one particle disappears during the time $\tau$, then 
$$\frac{d n}{d t} \approx \frac{\Delta n}{\tau} = \frac{\Delta N / V}{\tau} = \frac{1}{V \tau} .$$
Hence,
\begin{equation}\label{eq:tau}
\begin{gathered}
  \frac{1}{V \tau} \approx \frac{d n}{d t} = \frac12 \sum\limits_{i,j} \hat C_{i,j} n_i n_j = \frac12 \sum\limits_{i,j} \hat C_{i,j} N_i N_j / V^2, \hfill \\
  \tau = \frac{2V}{\sum\limits_{i,j} \hat C_{ij} N_i N_j} = \frac{2V \left( 1 - f_{ij} \right)}{\sum\limits_{i,j} C_{ij} N_i N_j},
\end{gathered}
\end{equation}
where $N_i$ is the number of particles of size $i$. Since Eq. \eqref{eq:tau} allows for aggregation of the particle with itself, we need to exclude such collisions, applying the rejection step with the probability $1/N_i$ whenever $i = j$. The coefficient $\frac12$ in Eq. \eqref{eq:tau} prevents from a double counting of each colliding pair.

If we fix $i$ and $j$ we can calculate average time between  collisions $\tau_{i,j}$ for some fixed pair of sizes. Similarly, taking into account  only aggregative collisions of particles of size $i$ and $j$, we can calculate the  average time between the according aggregation events, which we denote  $\tau_{i,j}^{agg}$:
\begin{equation}\label{eq:tauagg}
\begin{gathered}
  \frac{1}{V \tau_{i,j}^{agg}} \approx C_{i,j} n_i n_j = C_{i,j} N_i N_j / V^2, \hfill \\
  \tau_{i,j}^{agg} = \frac{V}{C_{ij} N_i N_j}.
\end{gathered}
\end{equation}

To determine the average change of temperatures during the aggregative collision of particles of size $i$ and $j$, we can use kernels $B_{ij}$ and $D_{ij}$. We find that
\begin{equation}\label{eq:deltat}
\begin{aligned}
  \frac{\Delta \left( n_{i+j} \theta_{i+j}^{agg} \right)}{\tau_{i,j}^{agg}} & = \frac{\left( N_{i + j} + 1 \right) \left( \theta_{i + j} + \Delta \theta_{i + j} \right) - N_{i+j} \theta_{i+j}^{agg}}{V \tau_{i,j}} \\
  & \approx B_{ij} N_i N_j / V^2, \\
  \frac{\Delta \left( n_{i} T_{i} \right)}{\tau_{i,j}^{agg}} & = \frac{\left( N_i - 1 \right) \left( \theta_i + \Delta \theta_i \right) - N_i \theta_i}{V \tau_{i,j}^{agg}} \\
  & \approx - D_{ij}^{agg} N_i N_j / V^2, \\
  \frac{\Delta \left( n_{j} T_{j} \right)}{\tau_{i,j}^{agg}} & = \frac{\left( N_j - 1 \right) \left( \theta_j + \Delta \theta_j \right) - N_j \theta_j}{V \tau_{i,j}^{agg}} \\
  & \approx - D_{ji}^{agg} N_i N_j / V^2,
\end{aligned}
\end{equation}
where we use only aggregative part  $D_{ij}^{agg}$ of the kernel $D_{ij}$, that defines temperature variation  during the aggregation events. Substitution $\tau_{i,j}^{agg}$ from Eq. \eqref{eq:tauagg} into Eq. \eqref{eq:deltat} yields for the temperature change:
\begin{equation}\label{aggtemp-eq}
\begin{gathered}
  \left\{ \begin{gathered}
  \Delta \theta_i = \frac{\theta_i - {D_{ij}^{agg}}/C_{ij}}{\left({{N_i} - 1}\right)} = - \frac{{\hat D_{ij}^{agg}}}{{{N_i} - 1}}, \hfill \\
  \Delta \theta_j = \frac{\theta_j - {D_{ji}^{agg}}/C_{ij}}{\left({{N_i} - 1}\right)} = - \frac{{\hat D_{ji}^{agg}}}{{{N_i} - 1}}, \hfill \\
  \Delta {\theta _k} = \frac{{{{B}_{ij}}/C_{ij} - {\theta _k}}}{{{N_k} + 1}} = \frac{{{{\hat B}_{ij}} - {\theta _k}}}{{{N_k} + 1}}, \hfill \\ 
\end{gathered}  \right. \hfill \\
  \hat D_{ij}^{agg} = \left( {\frac{{4(1 - g_{ij})}}{{3(1 - f_{ij})}} - 1} \right)\frac{{\theta _i^2}}{{{\theta _i} + {\theta _j}}}, \hfill \\
  {{\hat B}_{ij}} = \frac{{{\theta _i}{\theta _j} + \frac{{4(1 - g_{ij})}}{{3(1 - f_{ij})}} \cdot \frac{{{{\left( {i{\theta _i} - j{\theta _j}} \right)}^2}}}{{{{\left( {i + j} \right)}^2}}}}}{{{\theta _i} + {\theta _j}}}. \hfill \\ 
\end{gathered} 
\end{equation}

Similarly we can find the set of equations for the temperature variation  during the restitutive collisions using the other part $D_{ij}^{res} = D_{ij} - D_{ij}^{agg}$ of the kernel $D_{ij}$:
\begin{equation}\label{restemp-eq}
\begin{gathered}
  \left\{ \begin{gathered}
  \Delta {\theta _i} = - \frac{{\hat D_{ij}^{res}}}{{{N_i}}}, \hfill \\
  \Delta {\theta _j} = - \frac{{\hat D_{ji}^{res}}}{{{N_j}}}, \hfill \\ 
\end{gathered}  \right. \hfill \\
  \hat D_{ij}^{res} = \frac{{4g_{ij}}}{{3f_{ij}}} \cdot \frac{j}{{i + j}}\left( {1 + \varepsilon } \right)\left( {{\theta _i} - \frac{\left( {1 + \varepsilon } \right) j}{ 2 \left( i + j \right)}\left( {{\theta _i} + {\theta _j}} \right)} \right). \hfill \\ 
\end{gathered} 
\end{equation}

Note that with the above procedure, we can compute the average temperature change directly. This is significantly more efficient than applying the procedure of randomly choosing cluster speed from a Maxwellian velocity distribution. In other words, we effectively apply the mean-field approximation for temperatures. Also, note that in this version of MC, we use effective virtual particles corresponding to an ensemble of real particles. Hence we can straightforwardly double the system size any number of times.

If we now apply Gillespie \cite{gillespie} or inverse \cite{inverse} method, we need at least $O(M)$ operations after every collision, where $M$ is the maximum cluster size. Indeed, when $\theta_i$ changes, we need to recalculate $i$-th row and column of the collision kernel $\hat C$. However, if we use a low-rank approximation of the coagulation kernel \cite{MKSTB,matveev_CPC2018,meode}, only $O(r)$ elements under this approximation change; here $r$ is the rank of the system. 

Next, we give a brief description of the algorithm in the context of temperature-dependent ballistic aggregation. The paper with the detailed description of the low-rank Monte Carlo algorithm for the general case is in preparation.

To approximate the collision frequencies $\hat {C}_{ij}$ we use the following matrix $A$ of rank 3:
\begin{equation}\label{ac-eq}
\begin{aligned}
  & {A_{ij}} = \sqrt {\pi /2} {\left( {{i^{1/3}} + {j^{1/3}}} \right)^2}\sqrt {{\theta _i}}, \\
  & \frac{{{A_{ij}} + {A_{ji}}}}{{\sqrt 2 }} \leqslant {\hat C_{ij}} \leqslant {A_{ij}} + {A_{ji}}.
\end{aligned}
\end{equation}
Eq. (\ref{ac-eq}) shows that we can pick the probabilities up from the matrix $2A$ instead of $\hat C$ (we can replace $A_{ji}$ by $A_{ij}$ because of the collision symmetry). We use rejection sampling for the cases, where we have an overestimate, which happens with the probability $\frac{\sqrt{\theta_i + \theta_j}}{\sqrt{\theta_i} + \sqrt{\theta_j}} \leqslant 1 - 1/\sqrt{2} < 0.3$. This idea is similar to the majorant kernels approach exploited in Ref. \cite{majorant}. 

After the multiplication of the matrix $A$ by the vectors composed of $n_i$ and $n_j$, we still get the sum of 3 rank-1 matrices. Indeed, if
\[
  A = \sum\limits_{k = 1}^r {{u_{(k)}}v_{(k)}^T} ,\quad {u_{(k)}},{v_{(k)}} \in {\mathbb{R}^M},
\]
then $A_{ij} n_i n_j = \left(A \circ nn^T\right)_{ij}$, where $\circ$ denotes  the element-wise product. It can be also defined as
\[
A \circ nn^T = \sum\limits_{k = 1}^r \left( n u_{(k)} \right) \left( n v_{(k)} \right)^T.
\]

In order to keep track of the sums and quickly choose the sizes of the colliding particles, we construct a segment tree on each of the vectors $u_{(k)}$ and $v_{(k)}$. Then the update of the full structure requires $O(r \log M)$ operations and in case of ballistic agglomeration $r = 3$. In particular, segment trees contain the sums of all elements of $u_{(k)}$ and $v_{(k)}$ and thus we also know the sums of the $u_{(k)} v_{(k)}^T$ elements. To choose a pair of colliding particles, we firstly pick one of 3 rank-1 matrices with the probability, proportional to the sum of its elements ($O(r)$ operations) then the element of $u_{(k)}$ with the probability, proportional to its value ($O(\log M)$) and finally a column of $v_{(k)}$ (another $O(\log M)$ operations). Therefore, the total complexity of the method is $O(r \log M)$.

\section*{Derivation of the upper bounds $M_2(t)$ and $M_{2,\theta}(t)$}

Here we derive the differential equations {\bf (19)}  for the second moment of number density $M_2 = \sum_k  n_k k^2$ and the second moment of $n_k \theta_k$, equal to $M_{2,\theta} = \sum_k n_k \theta_k k^2$.

First, we estimate the upper bound of the sum  $\sum_k n_k T_k = n T_{avg}$, where $T_{avg}$ is the average temperature. Multiplying the system \eqref{nt-eq2} by $k$ and summing over all $k$ leads to
\[
\begin{gathered}
  \frac{d}{dt} n T_{avg} = \sum\limits_{i,j} { 2 \sqrt {2\pi } \sigma _{ij}^2\frac{{n_i}{n_j}}{{\sqrt {{\theta _i} + {\theta _j}} }} } \hfill \\
  \times \left( \frac{2}{3} (i + j){{\left( {\frac{{i{\theta _i} - j{\theta _j}}}{{i + j}}} \right)}^2}\left( {1 - {g_{ij}}} \right) \right. \hfill \\
  \left. - \frac{2}{3}\left(i\theta _i^2 + j\theta_j^2 \right)\left( {1 - {g_{ij}}} \right) \right. \hfill \\
  \left. { - \frac{4ij \left( \theta_i + \theta_j \right)}{3 \left({i + j}\right)}\left( {1 + \varepsilon } \right)\left( {{\theta _i} - \frac{j \left( 1 + \varepsilon \right)}{2 \left({i + j}\right)}\left( {{\theta _i} + {\theta _j}} \right)} \right)} g_{ij} \right). \hfill \\
\end{gathered}
\]
The sum of expressions containing $1 - g_{ij}$ is always negative. Bouncing term also cannot lead to increase of the kinetic energy, so it must be negative too. Therefore, the sum $\sum_k n_k T_k$ always decreases and thus can be bounded as 
\begin{equation}\label{eq:ntbound}
  n(t) T_{avg}(t) \leqslant n(0) T(0) = \operatorname{const}.
\end{equation}

Equation for the second moment $M_2(t)$ reads, 
\[
\begin{aligned}
  \frac{d M_2(t)}{dt} & = \sum\limits_{i,j} \left( \frac12 \left( i + j \right)^2 - i^2 \right) C_{ij} n_i n_j \\
  & = \sum\limits_{i,j} ij C_{ij} n_i n_j \\
  & \leqslant \sum\limits_{i,j} ij \sqrt{\pi / 2} \left( i^{1/3} + j^{1/3} \right)^2 \sqrt{\theta_i + \theta_j} n_i n_j \\
  & \leqslant \sum\limits_{i,j} ij \cdot 2 \sqrt{2 \pi} \left( i + j \right) \sqrt{\theta_i} n_i n_j \\
  & = 2 \sqrt{2 \pi} \sum\limits_i i \sqrt{\theta_i} n_i \sum\limits_j j^2 n_j \\
  & + 2 \sqrt{2 \pi} \sum\limits_i i^2 \sqrt{\theta_i} n_i \sum\limits_j j n_j \\
  & \leqslant 2 \sqrt{2 \pi M_1 n T_{avg}} M_2 + 2 \sqrt{2 \pi M_2 M_{2,\theta}} M_1 \\
  & \leqslant C_1 M_2 + C_2 \sqrt{M_2 M_{2,\theta}}.
\end{aligned}
\]
Here we used the Cauchy–Schwarz inequality $$\sum\limits_i n_i f(i) g(i) \leqslant \sqrt{\sum\limits_i n_i f(i) \sum\limits_j n_j g(j)}, $$
along with Eq. \eqref{eq:ntbound} to limit the sum $\sum_i n_i T_i$ and the equation for the total mass of the system $\sum_i n_i i = M_1 = \operatorname{const}$.

Now, we find the equation for the moment $M_{2,\theta}(t)$.
\[
\begin{gathered}
  \frac{d M_{2,\theta}(t)}{dt} = \sum\limits_{i,j} \left( \frac12 \left( i + j \right)^2 B_{ij} - i^2 D_{ij} \right) n_i n_j \hfill \\
  = \frac12 \sum\limits_{i,j} \left( \left( i + j \right)^2 B_{ij} - i^2 D_{ij} - j^2 D_{ji} \right) n_i n_j \hfill \\
  = \sqrt{2 \pi} \sum\limits_{i,j} \sigma_{ij}^2 \frac{1}{\sqrt{\theta_i + \theta_j}} \hfill \\
  \times \left( \left( i + j \right)^2 \theta_i \theta_j \left( 1 - f_{ij} \right) - \left( i^2 + j^2 \right) \theta_i \theta_j \left( 1 - f_{ij} \right) \right. \hfill \\
  + \left. \frac{4}{3} \left( i \theta_i - j \theta_j \right)^2 \left(1 - g_{ij} \right) - \frac{4}{3} \left( i^2 \theta_i^2 + j^2 \theta_j^2 \right) \left(1 - g_{ij} \right) \right. \hfill \\
  - \left. \frac{4ij \left( \theta_i + \theta_j \right)}{3 \left( i + j \right)} \left(1 + \varepsilon \right) \left(i \theta_i + j \theta_j \right. \right. \hfill \\
  - \left. \left. \left(1 + \varepsilon \right) \frac{ij}{i + j} \left( \theta_i + \theta_j \right) \right) g_{ij} \right) n_i n_j. \hfill
\end{gathered}
\]
Note, that the expression, containing $1 - g_{ij}$ is always negative and so can be neglected. Next, in the positive term of the bouncing part, we can replace $1 + \varepsilon$ by $2$, which will always increase the sum. Then
\begin{equation}\label{eq:m2tstart}
\begin{gathered}
  \frac{d M_{2,\theta}(t)}{dt} \leqslant \sqrt{2 \pi} \sum\limits_{i,j} \sigma_{ij}^2 \frac{1}{\sqrt{\theta_i + \theta_j}} \left( 2 i j \theta_i \theta_j \left( 1 - f_{ij} \right) \right. \hfill \\
  \left. - \frac{4ij}{3 \left( i + j \right)^2} \left(1 + \varepsilon \right) \left( \theta_i + \theta_j \right) \left( T_i - T_j \right) \left( i - j \right) g_{ij} \right) n_i n_j. \hfill
\end{gathered}
\end{equation}
Let us factor out $\left( T_i - T_j \right) \left( i - j \right) n_i n_j$ in the second (bouncing) term and denote by $E_{ij}$ the remaining product. Then
\begin{equation}\label{eq:cbounce}
  E_{ij} = \frac{4}{3} \sqrt{2 \pi} \sigma_{ij}^2 \sqrt{\theta_i + \theta_j}  \frac{ij}{\left( i + j \right)^2} \left(1 + \varepsilon \right) g_{ij}.
\end{equation}
Now we estimate the first term in \eqref{eq:m2tstart}. It can be bounded by the following expression:
\begin{eqnarray}
\label{eq:m2tfirst}
&&  2 \sqrt{2 \pi} \sum\limits_{i,j} \sigma_{ij}^2 \frac{1}{\sqrt{\theta_i + \theta_j}} i j \theta_i \theta_j \left( 1 - f_{ij} \right) n_i n_j \leqslant \hfill \\
&&  \leqslant \sqrt{2 \pi} \sum\limits_{i,j} \left( i + j \right) \frac{1}{\sqrt{\theta_i + \theta_j}} T_i T_j n_i n_j \hfill \nonumber \\
&&  = 2 \sqrt{2 \pi} \sum\limits_{i,j} \frac{i T_i T_j}{\sqrt{\theta_i + \theta_j}} n_i n_j \hfill \nonumber\\
&&  \leqslant 2 \sqrt{2 \pi} \sum\limits_{i,j} i \sqrt{i T_i} n_i T_j n_j \hfill \nonumber\\
&&  = 2 \sqrt{2 \pi} \sum\limits_i i \sqrt{i T_i} n_i \sum\limits_j T_j n_j \hfill \nonumber \\
&&  \leqslant 2 \sqrt{2 \pi} \sqrt{M_2 M_{2,\theta}} n(0) T(0) = C_3 \sqrt{M_2 M_{2,\theta}}. \hfill \nonumber
\end{eqnarray}
Substituting Eqs. \eqref{eq:cbounce} and \eqref{eq:m2tfirst} into \eqref{eq:m2tstart}, we arrive at
\[
  \frac{d M_{2, \theta}}{dt} \leqslant C_3 \sqrt{M_2 M_{2 \theta}} - \sum\limits_{i,j} \left( T_i - T_j \right) \left( i - j \right) E_{ij} n_i n_j,
\]
which is Eq. {\bf (19)} of the main text. 

\bibliography{PhaseDiagram.bbl}

\end{document}